\newif\ifaastex\aastexfalse

\ifaastex
\documentclass[manuscript]{aastex}
\else
\documentclass{emulateapj}
\fi

\ifaastex
\else
\let\jjdagger=\dagger
\let\jjddagger=\ddagger
\renewcommand{\dagger}{\ensuremath{\jjdagger}}
\renewcommand{\ddagger}{\ensuremath{\jjddagger}}
\fi

\usepackage{amsmath}

\newcommand{\gtwoninetytwo}{G292.0+1.8}
\newcommand{\cmthree}{\ensuremath{\mathrm{cm}^{-3}}}
\newcommand{\kms}{\ensuremath{\mathrm{km\,s}^{-1}}}
\newcommand{\msun}{\ensuremath{\mathrm{M}_{\odot}}}
\newcommand{\nh}{n_{\mathrm H}}

\shorttitle{Stellar Wind Interaction in SNR G292.0+1.8}
\shortauthors{Lee et al.}

\begin{document}




\title{The Outer Shock of the Oxygen-Rich Supernova Remnant
  G292.0+1.8: Evidence for the Interaction with
the Stellar Winds from its Massive Progenitor}

\author{Jae-Joon Lee\altaffilmark{1,2} Sangwook Park\altaffilmark{1},
{John P. Hughes\altaffilmark{3}},
{Patrick O. Slane\altaffilmark{4}},
{B. M. Gaensler\altaffilmark{5}},
{Parviz Ghavamian\altaffilmark{6}}
and {David N. Burrows\altaffilmark{1}}}

\altaffiltext{1}{Astronomy and Astrophysics Department, Pennsylvania
  State University, University Park, PA 16802}
\altaffiltext{2}{lee@astro.psu.edu}

\altaffiltext{3}{Department of Physics and Astronomy, Rutgers University, 136
Frelinghuysen Road, Piscataway, NJ 08854-8019}


\altaffiltext{4}{Harvard-Smithsonian Center for Astrophysics, 60 Garden Street,
Cambridge, MA 02138}

\altaffiltext{5}{Sydney Institute for Astronomy, School of Physics,
  The University of Sydney, NSW 2006, Australia}

\altaffiltext{6}{Space Telescope Science Institute, 3700 San Martin
  Drive, Baltimore, MD, 21218}



\begin{abstract}
  We study the outer-shock structure of
  the oxygen-rich supernova remnant G292.0+1.8, using a deep observation
  with the {\em Chandra X-ray Observatory}.  We measure
radial variations of the electron temperature and emission measure
that we identify as the outer shock
propagating into a medium with a radially
  decreasing density profile. The inferred ambient density structure
  is consistent with models for the circumstellar wind of a massive
  progenitor star rather than for a uniform interstellar medium.
  The estimated wind density ($n_{\mathrm{H}} = 0.1 \sim 0.3\
  \cmthree$) at the current outer radius ($\sim7.7$ pc) of the remnant
  is consistent with a slow wind from a red supergiant (RSG) star. 
  The
  total mass of the wind is estimated to be $\sim 15 - 40\, \msun$
  (depending on the estimated density range),
  assuming that the wind extended down to near the surface of the progenitor.
  The overall kinematics of \gtwoninetytwo\ are consistent with the
  remnant expanding through the RSG wind. 

\end{abstract}






\section{Introduction}


Studying the environments in which supernovae (SNe) explode and then
subsequently evolve is crucial to understand how SNe feed energy and
matter into the Galactic ecology.
This is particularly important for core-collapse
SNe which are believed to explode in a complicated environment
caused by the late stage evolution of a massive progenitor star.
The study of young core-collapse supernova remnants (SNRs)
interacting with their circumstellar media is important
in this regard.
The evolution of the SNR is not only governed by the
medium into which it is expanding but also by the details of its
explosion.  Therefore, the study of young core-collapse SNRs may also
provide valuable information about their explosion.



The environment of a massive star at the time of the explosion is
complicated by its evolutionary history.  Stars of initial mass $\lesssim
35\msun$ are likely to explode as red super giants (RSGs), while more
massive stars end as Wolf-Rayet (WR) stars with a possible earlier RSG
phase \citep{2005ApJ...619..839C}.  For a simplistic case where the
stellar wind had a constant velocity and a constant mass loss rate,
the circumstellar density profile of the freely flowing unshocked wind
is given by $\rho \propto r^{-2}$
where $\rho$ is the density of the wind at the distance $r$ from the
star. The structure and evolution of a SNR expanding in such a medium
differs  from that of a SNR expanding in a uniform density
medium.  \citet{1982ApJ...258..790C} derived a self-similar solution
for the 1-dimensional spherical expansion of SN ejecta whose density
profile is given as a power-law ($\rho \propto r^{-n}$) in a medium
with a radial density dependence ($\rho \propto r^{-s}$). The
evolution beyond the RSG wind
is more complicated (i.e., similarity solutions fail to apply) and
numerical studies are often required for a comprehensive description
\citep[e.g.,][]{2005ApJ...630..892D}.  A Wolf-Rayet phase expected in
more massive stars further complicates its environment
\citep[e.g.,][]{1989ApJ...344..332C,2007ApJ...667..226D}.



The circumstellar interaction of SNe has been
extensively studied, especially in the radio band
\citep[see][for a review]{2002ARA&A..40..387W}.  Such an interaction plays a
critical role in interpreting observations of $\gamma$-ray bursts
\citep[e.g.,][]{2006Natur.442.1008C}.  On the other hand,
observational studies of core-collapse SNRs expanding within a
circumstellar wind have not been well established so far.  Among more
than 260 Galactic SNRs, only a few known core-collapse SNRs show
circumstantial evidence of the shock--wind interaction \citep[][and
references therein]{2005ApJ...619..839C}, among which the most notable
example is Cas A.
The overall characteristics of Cas A are well described by a model of
a supernova interacting with a RSG wind with a mass loss rate of $\sim 2
\times 10^{-5} \msun$ yr$^{-1}$ for a wind velocity of 10 \kms
\citep{2003ApJ...593L..23C}.

Another well-known example of a young core-collapse SNR is G292.0+1.8.  \gtwoninetytwo\  is
one of the three known ``oxygen-rich'' SNRs
\citep{1979MNRAS.188..357G,1979MNRAS.189..501M} in the Galaxy, along
with Cas A and Puppis A. These SNRs show optical emission from fast-moving ($v
\gtrsim 1000\,\kms$) O-rich ejecta knots, which are suggested to be
synthesized
in the core of a massive star \citep[$>10$ \msun;
e.g.][]{2000ApJ...537..667B}.  The existence of the central pulsar
(PSR J1124-5916) and its wind nebula
\citep{2001ApJ...559L.153H,2002ApJ...567L..71C,2003ApJ...591L.139H,2003ApJ...594..326G}
clearly demonstrates that \gtwoninetytwo\ originated from a
core-collapse SN of a massive star.  A progenitor mass of $\sim 20 -
40$ \msun\ is suggested from the observed metal abundance structure of
an X-ray emitting ejecta
\citep{1994ApJ...422..126H,2003ApJ...583L..91G,2004ApJ...602L..33P}. Previous
X-ray studies showed an X-ray bright central belt-like structure,
which likely corresponds to shocked circumstellar medium (CSM)
\citep{2002ApJ...564L..39P}.
%
\citet{2002ApJ...564L..39P}
suggested that the belt traces the enhanced mass loss in the
progenitor star's equatorial plane.  \gtwoninetytwo\ also shows faint
X-ray emission along the outer boundary of the remnant, which
corresponds to a sharp-edged perimeter seen in the radio morphology
\citep{2003ApJ...594..326G}. This faint X-ray emission is produced by the
blast wave propagating into the ambient medium.
It shows a simple, nearly circular morphology all around the SNR
without showing significant substructures and is mostly free from
shocked metal-rich ejecta emission. Thus, this outer region of
\gtwoninetytwo\ is useful for the detailed study of the CSM and its
interaction with the SN blast wave.



In this paper, we present the results from our analysis of shocked CSM
in \gtwoninetytwo\ using our deep $\sim$500 ks {Chandra} observation
first presented by \cite{2007ApJ...670L.121P}, which provides an
excellent opportunity for the spectral study of this faint emission
from the outer boundary of the SNR.  We describe the {Chandra}
observation and the data reduction in
\S~\ref{sec:observ-data-reduct}. In
\S\S~\ref{sec:radial-profile-outer} and \ref{sec:azim-vari-asymm}, we
investigate radial and azimuthal variations of the X-ray spectral
properties of shocked CSM to show that the remnant is expanding in a
circumstellar wind. The implications of our results on the nature of
the circumstellar wind and the kinematics of the remnant are discussed
in \S~\ref{sec:discussion}.  \S~\ref{sec:summary-conclusion}
summarizes our results.

\section{Observations and Data Reduction}
\label{sec:observ-data-reduct}


We observed \gtwoninetytwo\
on 2006 September 13 -- October 20 \citep[][]{2007ApJ...670L.121P}
using the Advanced CCD Imaging Spectrometer (ACIS) on board
\emph{Chandra}. The entire remnant ($\sim9\arcmin$ in diameter) is covered
with ACIS-I ($17\arcmin \times 17\arcmin$ field of view) and the
pointing was selected to place the pulsar (PSR J1124-5916) near the
aimpoint.  The total effective exposure time is 506 ks after the data
reduction, with the exposure of individual observations ranging
between $\sim 40$ and $160$ ks (Table~\ref{tab:obssum}).
The level 1 event files were reprocessed to create new level 2 event
files. We applied parameters of the standard Chandra pipeline process,
except that we applied the VFAINT mode background cleaning and turned off
the pixel randomization. CIAO 3.4 and CALDB 3.2.3 were used for all the
reprocessing and analysis. We examined the overall background light curve
for periods of high background.
Only minor increases above the quiescent background level by a factor
of $\lesssim 2$
were noticed and their total duration is less than 1\% of the total
exposure. Therefore, we did not apply any light curve filtering.

We checked the relative astrometry between individual observations using
the position of the pulsar (PSR J1124-5916). We found less than
$0.1\arcsec$ deviations which are within typical on-axis astrometric
uncertainties of the ACIS.
For the purpose of the imaging analysis, event files were reprojected
to a common tangent point (that of ObsID 6679). We generated exposure
maps for each ObsID with the dead area
correction\footnote{\url{http://cxc.harvard.edu/ciao/why/acisdeadarea.html}}
applied. For the spectral analysis, spectrum and response files were
generated for individual ObsIDs and then merged, weighting by the
exposure time. Out-of-time events%
\footnote{\url{http://cxc.harvard.edu/proposer/POG\_CYC10/html/chap6.html\#sec:trailed-images}}
from the bright SNR emission along the CCD readout directions are
present, which in principle could affect our spectral analysis. For
the southeastern region, where the SNR emission is faint, the
estimated out-of-time event counts are 7\% of the total counts (source
and background).
However, the out-of-time events contribute uniformly along the CCD
readout direction and we find the local background subtraction
removes 
more than 90\% of out-of-time events in the source spectrum.  The
effects of out-of-time events to the background-subtracted spectrum are
thus negligible. 


\begin{deluxetable}{llrr}
\tablecolumns{4}
\tablewidth{0pc}
\tablecaption{Summary of Chandra observations\label{tab:obssum}}
\tablehead{
\colhead{ObsID} & \colhead{Date} & \colhead{Exposure} & \colhead{RollAngle\tablenotemark{\dagger}}\\
\colhead{} & \colhead{} & \colhead{[ks]} & \colhead{[$\arcdeg$]}
}
\startdata
{6680\tablenotemark{\ddagger}} & {9/13/2006} & {39} & {180}\\
{6678} & {10/2/2006} & {44} & {157}\\
{6679} & {10/3/2006} & {153} & {157}\\
{8447} & {10/7/2006} & {47} & {157}\\
{6677} & {10/16/2006} & {159} & {140}\\
{8221} & {10/20/2006} & {64} & {140}\\
\enddata
\tablenotetext{\dagger}{
    Roll angle describes the orientation of the Chandra instruments on the
    sky.
    }
\tablenotetext{\ddagger}{
     ObsID 6680 was conducted with 2 ACIS-S chips in addition to 4
     ACIS-I chips, and affected by telemetry saturation, reducing
     the net exposure time by $\sim25$\%. All later observations used only
     ACIS-I chips, and thus were not affected by the telemetry
     saturation.}
\end{deluxetable}

\section{Radial Profile of the Outer Shell of G292.0+1.8}
\label{sec:radial-profile-outer}

\subsection{Spectral Fitting}
\label{sec:observ-radi-prof}

\begin{figure}
  \plotone{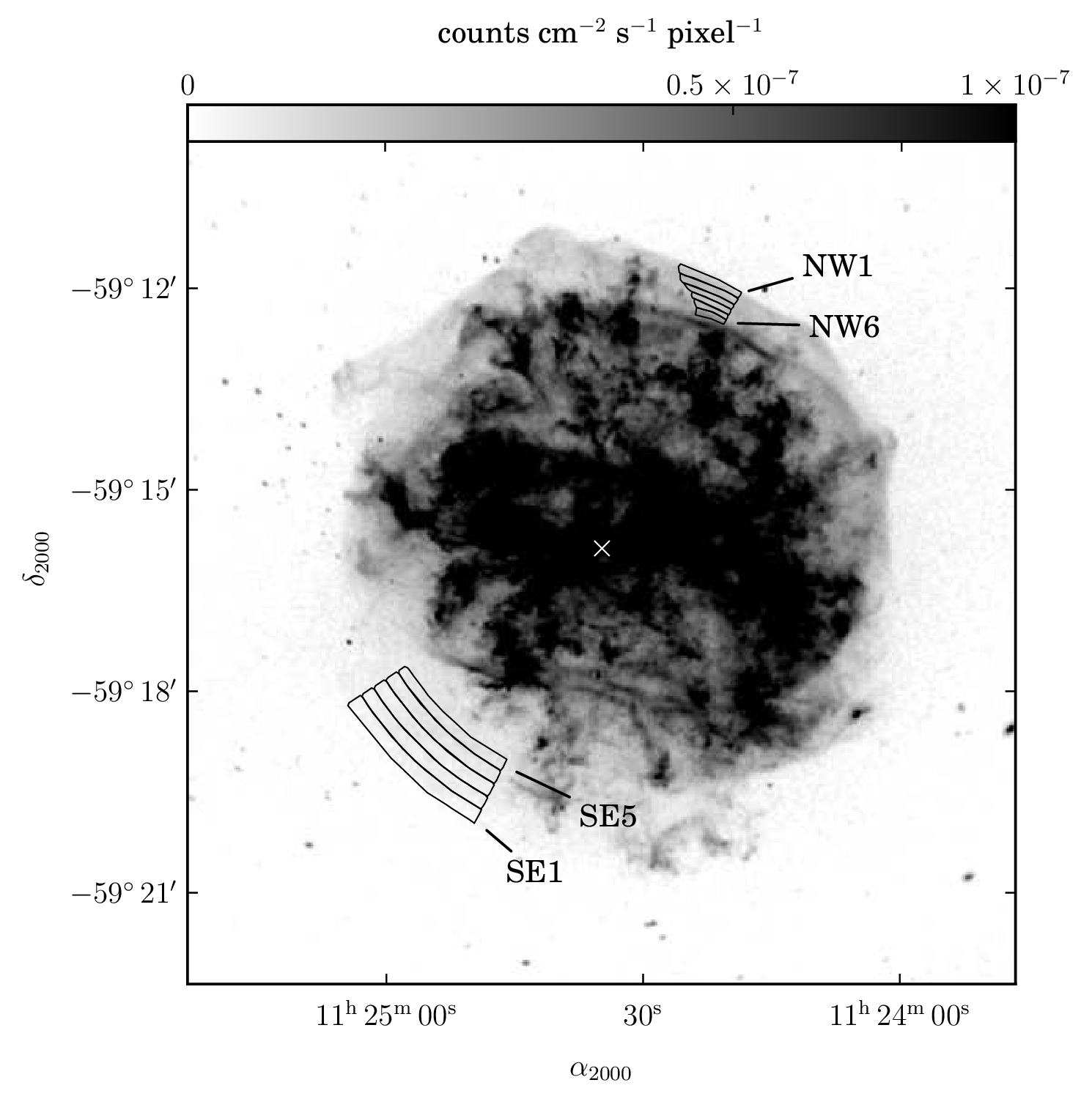}
  \caption{The $0.3-10$ keV band Chandra ACIS image of
    \gtwoninetytwo\ with regions used in the radial spectral analysis
    overlaid. The size of each individual region is adjusted to
    have a similar number of photon counts while avoiding any local
    features.
    The image is exposure-corrected and has a pixel size of $0\farcs492$.
    The grey scale is saturated in the bright filamentary
    structures near the central regions of the SNR to emphasize the
    faint outer shock.  
    The radio geometrical center of the SNR \citep{2003ApJ...594..326G}
    is marked as a white X.
   }
  \label{fig:region}
\end{figure}

The X-ray images of the deep Chandra observation clearly reveal the
faint outer blast wave all around the remnant (Fig.~\ref{fig:region}).
While the 
northern and western outer boundaries are well-defined and relatively
bright, that of the southeast is faint.  Although a slight
limb-brightening appears to be present in some parts of the
northwestern boundary, in general, the surface brightness of the outer
boundary does not show a significantly limb brightened
morphology.
The radial brightness variation is rather flat and even increasing
inward in some regions (e.g., southeast).
Given the core-collapse origin of the remnant, we consider
this overall morphology in the framework of the SNR
expanding in its circumstellar wind.

For a quantitative analysis, we investigate the radial variation of
the X-ray spectral characteristics near the outer boundary.
This approach is motivated by the fact that
the gas behind the forward shock preserves the information of the
medium it has swept up. 
\citet{1982ApJ...258..790C} found self-similar solutions for the
interaction of expanding stellar ejecta ($\rho \propto r^{-n}$, $n>5$)
with an external medium ($\rho \propto r^{-s}$): a remnant
expanding in a uniform ambient medium ($s=0$) would result in an
inward-decreasing density and an inward-increasing temperature
profile for the shocked ambient gas, while a remnant expanding in a medium with a wind profile
($s=2$) would show an inward-increasing density and an
inward-decreasing temperature.
Therefore, by studying the
current radial profile of the density and temperature, we can infer
whether the SNR is currently expanding into a density gradient.

\begin{figure}
  \ifaastex
  \epsscale{0.8}
  \else
  \epsscale{1.}
  \fi
  \plotone{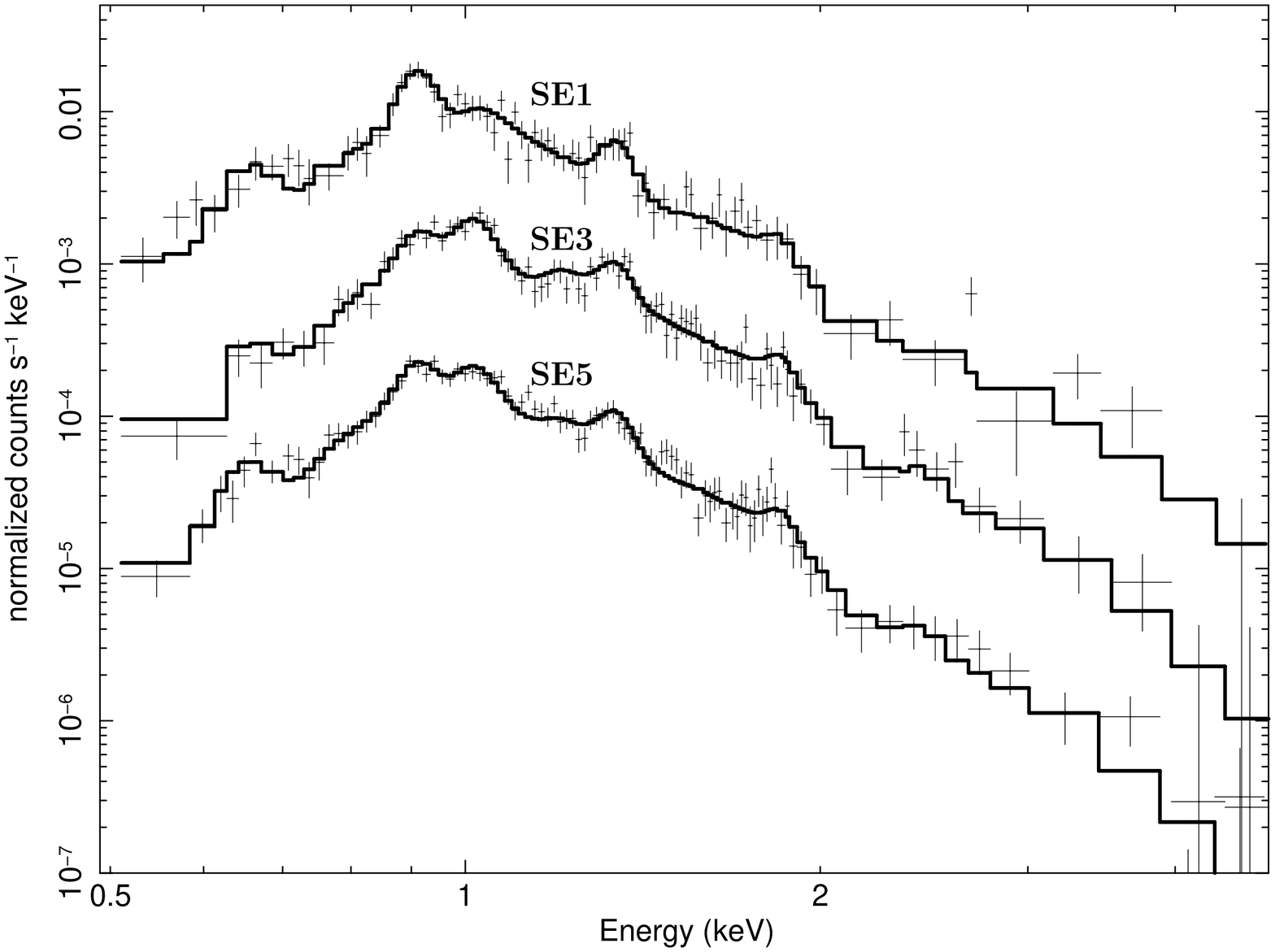}
  \caption{The Chandra ACIS spectra of the SE sector. Spectra at
    different radial distances (SE1, SE3, SE5) are shown. The
    spectra of SE3 and SE5 are arbitrarily scaled for the purpose of
    display.  The solid
    lines are the best-fit models.
  }
  \label{fig:spec}
\end{figure}

We selected two representative regions in
the northwest (NW) and southeast (SE). We examined various narrow band
images and X-ray color images of the
remnant 
and selected regions where the
contamination by ejecta emission is negligible and the area
is sufficiently large to study the radial spectral variation.
Each region (NW and SE) is divided into several radial
subregions to track the radial spectral variation
(Fig.~\ref{fig:region}). The photon statistics for each subregion are
$\sim3000-5000$ counts in the $0.5$-$8$ keV band.  The
extracted spectrum for each subregion is fitted with a single
nonequilibrium ionization (NEI) plane-parallel shock model
\citep[vpshock in Xspec v12,][]{2001ApJ...548..820B}.
%
The fits with variable metal abundances (O, Ne, Mg, Si, S and
Fe were fitted while the other species were fixed at their solar
values \citep{1989GeCoA..53..197A}) are statistically better
($\chi^2/\nu \sim 1.0$) than those with metal abundances frozen
($\chi^2/\nu \sim 1.2$) at solar values.
The fitted metal abundances
are generally subsolar, confirming that there is no significant ejecta contamination
in the selected regions. The derived spectral parameters from the
individual subregional fitting
are not well constrained due to the large
number of free parameters involved in the fit.
As the abundance pattern in each
radial region is similar within statistical uncertainties, we simultaneously fit all
subregions in each of the NW and SE regions with their metal abundances tied.
The variation of the absorbing column density when fitted independently is
statistically insignificant, and thus we tie the
absorbing column densities among subregions, as well.
We note that our fits result in a relatively higher $N_{\mathrm{H}}$ of
$\sim 5-6 \times 10^{21}
\mathrm{cm}^2$ than $N_{\mathrm{H}} \sim 3.2 \times 10^{21}
\mathrm{cm}^2$ from the spectral analysis of the pulsar
\citep{2001ApJ...559L.153H}.
Fixing $N_{\mathrm{H}}$ at $3.2 \times 10^{21}
\mathrm{cm}^2$ increases the temperature (by $\sim20$\%) and
decreases the normalization (by $\sim30$\%) in a systemic way and does
not fundamentally affect our results.
The above approach provides acceptable fits (example spectra and the
best--fit models are shown in Fig.~\ref{fig:spec}) for individual
subregions in each of the SE and NW regions with well constrained
spectral parameters.  The fitted parameters are consistent with those
from independent fits within statistical errors.
We utilize these fit
parameters for the following radial analysis.
The fit results for regions NW and SE are
summarized in Tables~\ref{tab:tied-fit} and \ref{tab:untied-fit}.


\subsection{Model Comparison}
\label{sec:model-comparison}

\begin{figure}
  \ifaastex
  \epsscale{.7}
  \fi
  \plotone{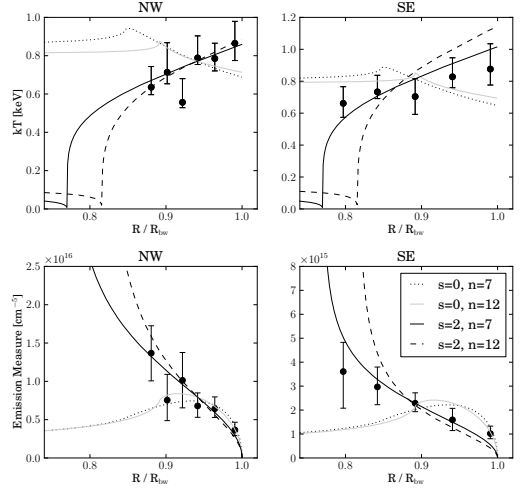}
  \epsscale{1.}
  \caption{Observed electron temperatures and emission measures,
    plotted as a function of the radial distance from the SNR center,
    for NW and SE regions. The radial distances are normalized by
    distance from the geometric center of the SNR to the blast wave
    ($R_{\mathrm bw}$) in each direction. Lines are model predictions
    from 
    \citeauthor{1982ApJ...258..790C} (\citeyear{1982ApJ...258..790C};
    see \S~\ref{sec:radial-profile-outer} for more details) for models
    of different set of parameters, where $s$ and $n$ are power-law
    index for the density profile of the ambient medium and of the
    expanding ejecta, respectively. Cusps in these lines mark the
    location of the contact discontinuity in the model.  The
    temperatures and emission measures of the model predictions have
    been scaled to match the observed values.}
  \label{fig:rad-prof}
\end{figure}

In Fig.~\ref{fig:rad-prof}, we plot the spectral variation as a
function of the radial distance of each subregion from the geometric
center of the remnant as determined by the radio continuum image
\citep[$\alpha_{2000}, \delta_{2000}$ = 11$^{\mathrm h}$ 24$^{\mathrm
  m}$ 34\fs8, -59\arcdeg 15\arcmin 52\farcs9,][]{2003ApJ...594..326G},
where the radial distance of each subregion is normalized by the
distance to the forward shock in that region.  As these regions are
located in the outer parts of the remnant, the uncertainty of the
remnant center does not make a significant effect on the results
presented in this section.  The radial variations of the spectral
parameters are consistent between the NW and SE regions
(Fig.~\ref{fig:rad-prof}).  The emission measure\footnote{ We use the
  conventional emission measure (EM = $\int n_e n_{\mathrm H} \,dl$)
  where the fitted volume emission measure ($\int n_e n_{\mathrm H}
  \,dV$) is normalized by the region area.} increases inward, while
the electron temperature does not.  The ionization timescale ($n_e t$)
shows a similar trend as the emission measure.

We overlay model predicted radial distributions of the temperature and
emission measure in Fig.~\ref{fig:rad-prof}, where the radial profiles
from \citet{1982ApJ...258..790C} are projected assuming a spherical
geometry.  \citet{1982ApJ...258..790C} provides models of density,
pressure and velocity of the gas as a function of the radial distance
from the center.  The model emission measure in
Fig.~\ref{fig:rad-prof} is calculated by integrating the density
square along the line of sight.  The temperature is not explicitly
provided in \citet{1982ApJ...258..790C}, and we have calculated it
from the thermal pressure and gas density.  The projected temperature
in Fig.~\ref{fig:rad-prof} is taken as an average temperature along
the line of sight weighted by density squared.  While we ignore the
variation in the intrinsic emissivity of the gas at different
temperatures, the variation is not significant ($<30$ \%) in the
temperature range of 0.6 -- 0.9 keV.  Models with four different sets
of $n$ and $s$ are plotted.
 The solutions of $s=2$ show distinguished radial variations
  from the solutions of $s=0$, which reflect their different radial
  trends in density and temperature as briefly described in
  \S~\ref{sec:observ-radi-prof}.  The projected temperature of the
  shocked ambient medium of $s=2$ solutions shows a decrease toward
  the contact discontinuity (CD), in contrast to the increase in $s=0$
  solutions.  The emission measure of $s=2$ solutions monotonically
  increases toward the CD. The emission measure of $s=0$ solutions
  initially increases due to the geometrical effect, and then starts
  to decrease. These overall trends of the emission measure and
  temperature profiles are dominated by the ambient density structure
  ($s = 0$ vs.~$2$) rather than by the ejecta profile. The primary
  effect by the power law index $n$ of the ejecta is on the distance
  between the forward shock front and the CD: e.g., a steeper ejecta
  profile makes the CD close to the forward shock.
Fig.~\ref{fig:rad-prof} demonstrates that the general tendency of the
observed temperature and emission measure are well described by models
with $s=2$ (the solid and dashed lines). The model of $s=2$, $n=7$, in
particular, is in good agreement with the observations, where our fits
show reduced $\chi^2$ of $0.5 \sim 1$ in both SE and NW regions. On
the other hand, $s=0$ models (the gray and the dotted lines) show
significant discrepancies especially in the SE region with a reduced
$\chi^2$ of $\sim 3$.
The observed radial structures of the temperature and the emission measure
are inconsistent with 
the remnant expanding in a uniform medium where the postshock inner
region is expected to result in a lower density
and a higher temperature than the region immediately behind the
blast wave \citep{1982ApJ...258..790C}.



We note that the temperature from the X-ray spectral fit is the
electron temperature, while the temperature from the model is the
ion temperature. These two can differ considerably in
non-radiative shocks faster than 1000 \kms\ \citep[e.g.,][and
references therein]{2007ApJ...654L..69G}.
Assuming the SNR age of 3000 yr \citep[][this age is adopted
throughout this paper]{2005ApJ...635..365G,2009ApJ...692.1489W}, the
average blast wave velocity would be $2500\,\kms$ for a
remnant radius of $\sim$ 7.7 pc (see \S~\ref{sec:sn-expl}), and the
blast wave should have been faster in the past.  The observed ratio
between the electron temperature ($\mathrm{T}_{\mathrm{e}}$) and the
proton temperature ($\mathrm{T}_{\mathrm{p}}$) immediately behind the
shock is known to vary with shock velocities
\citep{2007ApJ...654L..69G}.
But the detailed variation of
$\mathrm{T}_{\mathrm{e}}/\mathrm{T}_{\mathrm{p}}$ in the shocks of
$v_s \gtrsim 2500\,\kms$ is unclear and we assume a constant
value of $\mathrm{T}_{\mathrm{e}}/\mathrm{T}_{\mathrm{p}}$
across the radius of the postshock regions used in our analysis
(the actual value does not matter for our analysis).
We note that Coulomb equilibration in the downstream region
will eventually equilibrate ions and electrons, and the electron
temperature will increase.
This would increase the downstream electron temperature.
Qualitatively, the modeled radial profile of the electron temperature
would thus be flatter (in case of the $s=2$ models) than that of the ion
temperature, which would still be consistent with our data.




The analytic model of \citet{1982ApJ...258..790C} is a similarity
solution assuming a power law distribution of the ambient density and
ejecta, which may not be entirely adequate for actual SNRs.  For
example, the $s=2$ models have the initial wind density profile of
infinite at $r=0$ and predict an infinite density and zero temperature
at the CD.  However, we believe that this
solution is a good approximation to the real case for the regions away
from the CD.  While the density distribution of the ejected material
depends on the structure of the progenitor star, a power law envelope
surrounding a flat (or shallower) core is a reasonable
approximation 
\citep{1989ApJ...341..867C,1999ApJ...510..379M}.  
 The assumption of the power law density of ejecta will
  eventually break down as the reverse shock reaches the core.  While
  this may significantly modify the structure of the shocked ejecta
  material, the effect on the structure of the shocked ambient gas would
  not be significant.  The \citet{1982ApJ...258..790C} models do not
  include the effects of cosmic ray acceleration in the shock. While
  the efficient cosmic ray acceleration may modify the underlying
  shock structure, the overall trend of the radial variation is likely to
  be still valid \citep[e.g.,][]{2000ApJ...543L..57D}.
  Therefore, we conclude that the observed trend of the decreasing
  temperature and increasing emission measure of the shocked ambient
  gas toward the CD is a robust indicator that the blast wave is
  currently expanding into a radially-decreasing density profile
  ($s=2$) rather than a constant density($s=0$).
%
  This result provides direct observational evidence that the blast
  wave of \gtwoninetytwo\ is currently propagating through the stellar
  winds produced by its massive progenitor.  On the other hand, our
  results may not constrain the power law index of the ejecta
  ($n$). The apparent effects of $n$ is the change in the relative
  distance from the CD to the outer boundary. However, this distance
  can be varied by other physical processes which are not properly
  included in the model, e.g., cosmic ray accelerations. Also, the
  propagation of the reverse shock into the shallow ejecta core may
  result in the decrease of the effective power-law index.  Therefore,
  while the observation shows better agreement with the model of $n=7$
  than the model of $n=12$, this may not be interpreted in favor of
  the $n=7$ model over the $n=12$ model.



\newcommand{\TableRadTiedCaption}{Fitted $N_{\mathrm H}$ and metal
  abundances
  for the regions used for the radial analysis.
  \label{tab:tied-fit}}

\newcommand{\TableRadTiedNote}{These parameters are fitted after being
  tied among subregions. Results for 
  spectral parameters varied freely are presented in
  Table~\ref{tab:untied-fit}. Metal abundances are with respect to
  solar \citep{1989GeCoA..53..197A}. Throughout the paper, errors are
  shown in 90\% confidence range unless explicitly specified
  otherwise.}

\begin{deluxetable*}{cccccccc}
\tablecolumns{8}
\tablewidth{0pc}
\tablecaption{\TableRadTiedCaption}
\tablehead{
\colhead{Sector} & \colhead{$N_\mathrm{H}$ [$10^{21}$ cm$^{-2}$]} & \colhead{O} & \colhead{Ne} & \colhead{Mg} & \colhead{Si} & \colhead{S} & \colhead{Fe}
}
\startdata
{NW} & {5.8$_{-0.8}^{+0.4}$} & { 0.67$_{- 0.10}^{+ 0.19}$} & { 0.75$_{- 0.07}^{+ 0.07}$} & { 0.35$_{- 0.06}^{+ 0.09}$} & { 0.36$_{- 0.07}^{+ 0.07}$} & { 0.63$_{- 0.24}^{+ 0.25}$} & { 0.24$_{- 0.05}^{+ 0.06}$}\\
{SE} & {5.0$_{-0.8}^{+0.3}$} & { 0.36$_{- 0.10}^{+ 0.13}$} & { 0.64$_{- 0.11}^{+ 0.12}$} & { 0.21$_{- 0.03}^{+ 0.04}$} & { 0.13$_{- 0.03}^{+ 0.03}$} & { 0.12$_{- 0.11}^{+ 0.11}$} & { 0.07$_{- 0.02}^{+ 0.03}$}\\
\enddata
\tablecomments{\TableRadTiedNote}
\end{deluxetable*}

\newcommand{\TableRadUntiedCaption}{Spectral parameters for NW and SE
  regions.\label{tab:untied-fit}}

\newcommand{\TableRadUntiedCommentsA}{Regions are listed on the
  decreasing order of their distance from the center; i.e., the top
  row in each region represents the outermost region. Based on the
  simultaneous fit for each region, $\chi^2/\nu = {3039}/{3042}$ and
  ${2415}/{2534}$ for NW and SE regions, respectively.}

\begin{deluxetable}{ccccc}
\tablecolumns{5}
\tablewidth{0pc}
\tablecaption{\TableRadUntiedCaption}
\tablehead{
\colhead{Region} & \colhead{Distance} & \colhead{$kT_e$} & \colhead{$\log n_e t$} & \colhead{E.M.}\\
\colhead{} & \colhead{[$\arcmin$]} & \colhead{[keV]} & \colhead{[cm$^{-3}$s]} & \colhead{[$10^{16}$ cm$^{-5}$]}
}
\startdata
{NW1} & {4.33} & { 0.86$_{- 0.09}^{+ 0.11}$} & { 10.41$_{- 0.06}^{+ 0.10}$} & { 0.37$_{- 0.07}^{+ 0.10}$}\\
{NW2} & {4.21} & { 0.79$_{- 0.06}^{+ 0.08}$} & { 10.74$_{- 0.13}^{+ 0.14}$} & { 0.64$_{- 0.11}^{+ 0.15}$}\\
{NW3} & {4.11} & { 0.79$_{- 0.03}^{+ 0.11}$} & { 10.75$_{- 0.13}^{+ 0.10}$} & { 0.68$_{- 0.15}^{+ 0.17}$}\\
{NW4} & {4.03} & { 0.56$_{- 0.03}^{+ 0.12}$} & { 11.34$_{- 0.13}^{+ 0.18}$} & { 1.01$_{- 0.36}^{+ 0.36}$}\\
{NW5} & {3.94} & { 0.71$_{- 0.06}^{+ 0.15}$} & { 10.93$_{- 0.20}^{+ 0.30}$} & { 0.75$_{- 0.27}^{+ 0.34}$}\\
{NW6} & {3.85} & { 0.64$_{- 0.04}^{+ 0.11}$} & { 11.05$_{- 0.19}^{+ 0.13}$} & { 1.37$_{- 0.36}^{+ 0.36}$}\\
\sidehead{}

{SE1} & {4.30} & { 0.88$_{- 0.10}^{+ 0.16}$} & { 10.69$_{- 0.16}^{+ 0.13}$} & { 0.10$_{- 0.02}^{+ 0.03}$}\\
{SE2} & {4.08} & { 0.83$_{- 0.07}^{+ 0.12}$} & { 11.13$_{- 0.10}^{+ 0.15}$} & { 0.16$_{- 0.04}^{+ 0.05}$}\\
{SE3} & {3.87} & { 0.70$_{- 0.11}^{+ 0.11}$} & { 11.51$_{- 0.08}^{+ 0.07}$} & { 0.23$_{- 0.04}^{+ 0.04}$}\\
{SE4} & {3.65} & { 0.73$_{- 0.04}^{+ 0.10}$} & { 11.37$_{- 0.15}^{+ 0.17}$} & { 0.30$_{- 0.07}^{+ 0.08}$}\\
{SE5} & {3.46} & { 0.66$_{- 0.09}^{+ 0.10}$} & { 11.32$_{- 0.18}^{+ 0.21}$} & { 0.36$_{- 0.15}^{+ 0.12}$}\\
\enddata
\tablecomments{\TableRadUntiedCommentsA}
\end{deluxetable}

\section{Azimuthal Variation}
\label{sec:azim-vari-asymm}

\begin{figure}
  \plotone{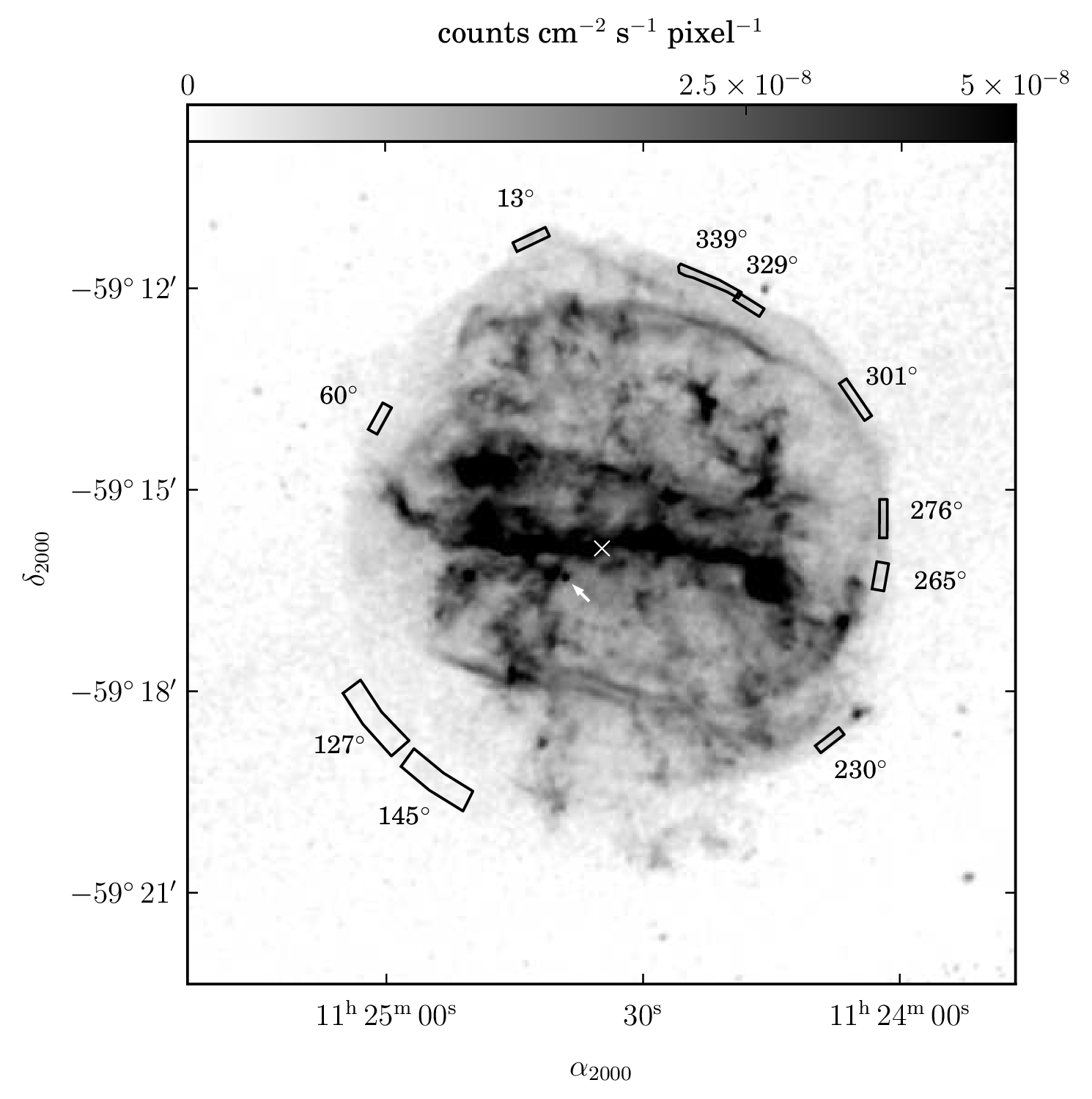}
  \caption{
    The $0.7-0.84$ keV band Chandra ACIS image of
    \gtwoninetytwo. The energy band has been selected to emphasize the
    equatorial belt.
    The regions along
    the remnant's outer boundary used for the azimuthal analysis are
    marked as black boxes with the corresponding position angles
    (north through east) indicated.
    The radio geometrical center of the SNR \citep{2003ApJ...594..326G}
    is marked as a white X.
    The position of the pulsar J1124-5916 is indicated by a white
    arrow.
  }
  \label{fig:wiind-center}
\end{figure}

Fig.~\ref{fig:wiind-center} shows regions we chose for the azimuthal
study. We selected narrow regions (thickness $10\arcsec \sim 20
\arcsec$) along the outer boundary of the remnant.
We avoided regions in which the contamination from the ejecta emission
is significant
We extracted the spectrum from each region and fitted it
independently with the {\it vpshock} model.
The photon statistics in these regions are a few thousand counts which are
similar to those in subregions used in the radial analysis. The fitted
spectral parameters are poorly constrained when the metal abundances are
freely varied,
and we fix the metal abundances to the average values as obtained from the
radial analysis (NW and SE regions). The fitted absorbing column
density varies between $5.5$ and $7.0\times 10^{21}$ cm$^{-2}$.
However, the $N_H$ variation is statistically insignificant, e.g., the
fits do not significantly change when $N_H$ is fixed at the mean
value. Thus, in the following discussion, we use results from the fits
where we fix $N_H$ at their mean of $6.2 \times 10^{21}$~cm$^{-2}$.


\newcommand{\TableAzCaption}{X-ray spectral parameters of the regions
  used in the azimuthal analysis. \label{tab:az-fit}}
\newcommand{\PaNote}{Position angles to each
  region are measured east of north from the geometrical center.}
\newcommand{\DofNote}{For all individual spectra, the degree of
  freedom ($\nu$) is 372.}
\newcommand{\TableAzComments}{\DofNote}
\begin{deluxetable}{ccccc}
\tablecolumns{5}
\tablewidth{0pc}
\tablecaption{\TableAzCaption}
\tablehead{
\colhead{P.A.\tablenotemark{\dagger}} & \colhead{$kT_e$} & \colhead{$n_{\mathrm{H}}$} & \colhead{$\log n_e t$} & \colhead{$\chi^2/\nu$\tablenotemark{\ddagger}}\\
\colhead{[$^{\circ}$]} & \colhead{[keV]} & \colhead{[cm$^{-3}$]} & \colhead{[$10^{10}$ cm$^{-3}$s]} & \colhead{}
}
\startdata
{ 13} & { 0.86$_{- 0.07}^{+ 0.07}$} & { 1.06$_{- 0.06}^{+ 0.08}$} & {10.36$_{-0.12}^{+0.15}$} & {1.06}\\
{ 59} & { 0.76$_{- 0.22}^{+ 0.17}$} & { 0.90$_{- 0.08}^{+ 0.27}$} & {10.79$_{-0.35}^{+0.61}$} & {0.89}\\
{126} & { 0.71$_{- 0.10}^{+ 0.09}$} & { 0.51$_{- 0.03}^{+ 0.05}$} & {10.81$_{-0.21}^{+0.24}$} & {0.97}\\
{144} & { 0.73$_{- 0.06}^{+ 0.07}$} & { 0.58$_{- 0.04}^{+ 0.05}$} & {10.43$_{-0.15}^{+0.20}$} & {1.05}\\
{229} & { 0.84$_{- 0.09}^{+ 0.10}$} & { 1.07$_{- 0.06}^{+ 0.07}$} & {10.68$_{-0.19}^{+0.21}$} & {0.90}\\
{264} & { 0.80$_{- 0.11}^{+ 0.12}$} & { 0.96$_{- 0.06}^{+ 0.09}$} & {10.77$_{-0.24}^{+0.21}$} & {0.92}\\
{275} & { 0.69$_{- 0.06}^{+ 0.06}$} & { 1.43$_{- 0.10}^{+ 0.14}$} & {10.40$_{-0.16}^{+0.18}$} & {0.93}\\
{300} & { 0.66$_{- 0.12}^{+ 0.12}$} & { 0.85$_{- 0.11}^{+ 0.20}$} & {10.30$_{-0.22}^{+0.43}$} & {0.90}\\
{329} & { 0.84$_{- 0.09}^{+ 0.09}$} & { 1.22$_{- 0.07}^{+ 0.10}$} & {10.48$_{-0.20}^{+0.21}$} & {0.95}\\
{338} & { 0.68$_{- 0.05}^{+ 0.05}$} & { 1.21$_{- 0.07}^{+ 0.08}$} & {10.41$_{-0.11}^{+0.13}$} & {1.01}\\
\enddata
\tablenotetext{\dagger}{\PaNote}
\tablenotetext{\ddagger}{\DofNote}
\end{deluxetable}

\begin{figure}
  \ifaastex
  \epsscale{0.6}
  \else
  \epsscale{0.8}
  \fi
  \plotone{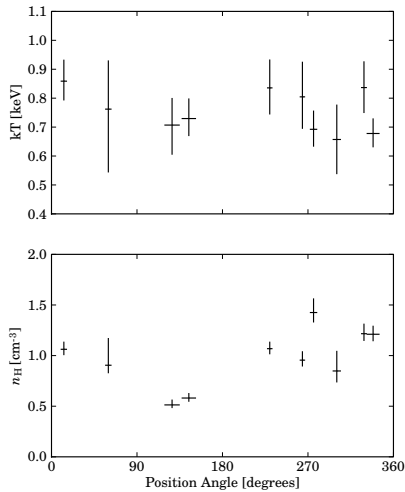}
  \caption{The azimuthal variation of the electron temperature and the
    postshock hydrogen density. Position angles are measured from the
    geometrical center as counter-clockwise from the north. The error
    bars in the position angle represent the angular extent of each
    region in Fig.~\ref{fig:wiind-center}. }
  \label{fig:az-vari}
  \epsscale{1.}
\end{figure}


The azimuthal variations of the electron temperature and the hydrogen
number density are presented in Table~\ref{tab:az-fit} and
Fig.~\ref{fig:az-vari}. The densities are estimated from the volume
emission measures. The volume corresponding to each region is calculated by
pixel-by-pixel integration of the path-lengths over the projected
area. In this calculation, we assumed a sphere centered at the
geometrical center of the remnant at a distance of 6 kpc
\citep{2003ApJ...594..326G}. The radius of the sphere is separately
determined for each region as the projected distance of the farthest
(from the geometrical center of the remnant) pixel in the region.
The uncertainties on the electron temperature are too large
to constrain any
possible variations
around the SNR boundary.  On the other hand, in the SE regions (PA $\sim
120\arcdeg - 150\arcdeg$), the density shows a significant drop by a
factor 2 from the average ($\sim\ 1\,\cmthree$).
The gas density shown in Fig.~\ref{fig:az-vari} is the postshock hydrogen
number density.  Assuming a compression factor of 4 from a strong gas
shock, this corresponds to a preshock gas density of $\sim\ 0.1\
\cmthree$ toward the SE and $0.2-0.3\,\cmthree$ in other regions.


\section{Discussion}
\label{sec:discussion} {Our observation provides direct
  evidence of \gtwoninetytwo\ interacting with its CSM wind, which is
  consistent with the core-collapse origin of the remnant (\S
  1). Our results also can help reveal the nature of the progenitor
  star by the observed wind properties.  In the followings, we discuss
  the nature of the wind with which \gtwoninetytwo\ is currently
  interacting and the overall kinematics of the remnant. We also
  briefly discuss the possible evolutionary paths of the progenitor
  star.

\subsection{The Progenitor's Stellar Wind}
\label{sec:nature-wind-expl}


\label{sec:wind-mass}

The expansion of SN blast waves within circumstellar winds has
been extensively studied with SNe observations. The shocked CSM emits in
radio continuum and also in X-rays. The mass loss rates of the
progenitor stars, estimated by modeling their light curve, are
$\sim10^{-6}\,\msun\, \mathrm{yr}^{-1}$ for type Ib/c and $\sim 10^{-4} -
10^{-5}\,\msun\, \mathrm{yr}^{-1}$ for type II SNe
\citep{2003LNP...598..145S}. The circumstellar interaction of SNRs, on
the other hand, has not been well established, although there has been
some circumstantial evidence \citep{2005ApJ...619..839C}.

The radial profiles of the postshock temperature and density in
\gtwoninetytwo\ argue that the remnant is expanding within
the circumstellar wind. A slight limb brightening is seen in a few
regions in NW, and we attribute it to a local effect, e.g., the shock
is probably encountering locally dense regions in the wind.
The overall dynamics of the remnant would not be significantly altered
by such a small local effect, and we consider that the SNR shock has
not yet reached the termination shock of the wind.
The radial profile of the wind density with a constant mass loss rate
of $\dot{M}$ and a wind velocity of $v_w$ is given by $\rho =
{\dot{M}} / {4 \pi r^2 v_w}$.  Adopting the geometrical center of
the remnant as the wind center, where we estimate an average shock
radius of $r_b = 7.7\, d_6$ pc, 
we estimate
\[\dot{M} = 2 \times 10^{-5}
\left(\frac{v_w}{10\,\kms}\right)
\left(\frac{\nh}{0.1\, \cmthree}\right)
\left(\frac{r}{7.7 \mathrm{pc}}\right)^2\,\msun\, \mathrm{yr}^{-1}.
\]
A slow wind velocity ($v_w \sim 10\, \kms$) appropriate for the RSG
winds implies a mass loss rate
$\dot{M} = 2-5 \times 10^{-5}\,\msun\, \mathrm{yr}^{-1}$ for the observed
density range ($\nh = 0.1 - 0.3\, \cmthree$), which is
consistent with the observations of type II SNe. 
On the other hand, assuming a fast ($v_w >
1000\, \kms$) Wolf-Rayet wind results in $\dot{M} > 10^{-3}\,\msun\,
\mathrm{yr}^{-1}$, a few orders of magnitudes larger than what is
inferred from type Ib/c SNe observations.

The integrated mass of the wind swept up by the remnant,
assuming the wind extends down to $r=0$,  is
\[M_w = 15 \left(\frac{\nh}{0.1\, \cmthree}\right)
\left(\frac{r_b}{7.7 \mathrm{pc}} \right)^3 \,\msun. \]
Accounting for the azimuthal density variation in the
  postshock regions, we estimate the mass range of $M_w \sim 15 -
  40\,\msun$.  There are further uncertainties involved in this
  estimate due to the various assumptions we made.
  Our density measurement scales as inverse square root of the assumed
  distance to the remnant ($\rho \propto d^{-1/2}$), thus $M_w \propto
  d^{5/2}$. 
  For example, a slightly smaller distance of 5 kpc
  \citep[which is within uncertainties in the distance measurements to
  \gtwoninetytwo,][]{2003ApJ...594..326G} will reduce the mass
  estimate by $\sim 40\%$.  Another source of uncertainties could be
  our simple assumption on the remnant geometry, a sphere centered at
  the geometric center of the remnant.  If the wind is clumped, a
  local density fluctuation may have led us to overestimate the
  average density, as the observed emission measure is likely biased
  by the high density regions.  The mass estimate will also be reduced
  if the shock compression ratio is higher than 4 due to efficient CR
  acceleration, while this effect may not be significant as there is
  no evidence for nonthermal X-ray emission in \gtwoninetytwo.  We
  assume that the mass loss remained constant until the moment of the
  explosion. However, observations of some SNe suggest that their mass
  loss rates have decreased with time before the explosion
  \citep[e.g.,][]{1994ApJ...432L.115V}.  Also, the progenitor star
  might have evolved through different stages (such as Blue Super
  Giant or Wolf-Rayet star).

  Theoretical studies of the stellar evolution suggests that stars of
  less massive than 35~\msun\ end their life during the RSG phase,
  while more massive stars evolve to WR phase with a possible earlier
  RSG phase. Given that the wind is likely a dense RSG wind, the progenitor
  star of \gtwoninetytwo\ might have a initial mass less massive than
  35~\msun.  In this regards, the wind mass as large as $40\,\msun$
  may not be realistic for stellar evolutionary models.  However, we note
  that there could be uncertainties in the modeling of the late stage
  of the stellar evolution, which is very sensitive to mass loss rate of
  the star, as the stellar mass loss rate is rather poorly constrained
  by observations.  On the other hand, previous nucleosynthesis
  studies of G292.0+1.8 using X-ray emission from metal-rich ejecta
  suggested a progenitor mass of $20-40\,\msun$
  \citep{1994ApJ...422..126H,2003ApJ...583L..91G,2004ApJ...602L..33P},
  while these progenitor mass estimates were limited by nondetection
  of the explosive nucleosynthesis products in this SNR
  \citep{2004ApJ...602L..33P}.  Considering that there are large
  uncertainties on both estimates of the wind and the progenitor
  masses, our estimate of the wind mass is in plausible agreement with
  a general picture: the progenitor of G292.0+1.8 was a massive
star which lost a significant amount of its mass during the
pre-SN evolution.

\label{sec:wind-center}

Our azimuthal analysis in \S~\ref{sec:azim-vari-asymm}
suggests an azimuthal density variation. Since the outer boundary
of the remnant has unlikely reached the termination shock of the RSG
wind, the observed density variation would represent the variation in
the wind density. The wind density in the SE boundary of the remnant
is smaller than those in other regions by a factor of $\sim 2$. 
While this could be due to some local effects,
the wind might have been asymmetric. 
Alternatively,
the center of the wind might have not been at the geometric center of
the SNR.  The asymmetric SN explosion might have caused the southeastern part
of the remnant to expand more rapidly than in NW, and the azimuthal density
variation could be due to an offset of the progenitor star from the
SNR's geometric center.
Recently, \citet{2009ApJ...692.1489W} studied proper motion of
[\ion{O}{3}] ejecta and located the estimated expansion center of
ejecta knots, which turned out to be consistent with the SNR's
geometric center.
However, we note that the observed proper motions of many individual
optical knots show significant deviations from the best-fit model
which indicated the explosion center at the geometric center of the
SNR. Thus, the origin of the azimuthal density variation along the
outer boundary of G292.0+1.8 is not conclusive.


\subsection{Supernova Remnant Kinematics}

\label{sec:sn-expl}

The morphology of the remnant, especially when it is young, traces its
explosion characteristics.  The kinematic properties of \gtwoninetytwo, such
as the age and the explosion energy, have been estimated in the previous
works. However, \gtwoninetytwo\  was previously assumed to be expanding in a
uniform medium instead of the circumstellar wind
\citep[e.g.,][]{2003ApJ...583L..91G,2003ApJ...594..326G}, although \gtwoninetytwo\
is believed to be a core-collapse SNR.
In fact, detailed observational studies of young core-collapse SNRs
accounting for the evolution in a circumstellar wind material have
been conducted for only a limited number of remnants such as Cas A
\citep{2003ApJ...593L..23C,2003ApJ...597..347L}.
Here, we estimate kinematic
properties of \gtwoninetytwo, explicitly accounting for its evolution in the
circumstellar wind.

For simplicity, we assume a spherically symmetric explosion.  Like the
SNR evolution in a homogeneous medium, the initial evolution of the
remnant is dominated by the characteristics of the ejecta (the
ejecta-dominated phase) and makes a transition to the Sedov-Taylor
(ST) phase as the swept-up mass becomes larger than the ejecta
mass. \citet[][TM99 hereafter]{1999ApJS..120..299T} discussed a
unified solution which describes a spherically symmetric interaction
of the ejected stellar material, which has a flat core surrounded by a
power-law envelope, with an ambient medium.  TM99 focused on the
evolution in the homogeneous medium. \citet[][LH03
hereafter]{2003ApJ...597..347L} extended TM99's work to evolution
in a stellar wind.  The characteristic age\,($t_0$) and size\,($x_0$)
of the remnant, which mark the approximate transition from the
ejecta-dominated (ED) phase to the ST phase, are given by
\[
t_0 = 5633 M_{ej}^{3/2} E_{51}^{-1/2} (\nh R_b^2)^{-1} \mathrm{yr}
\]
\[
x_0 = 40.74 M_{ej} (\nh R_b^2)^{-1} \mathrm{pc}
\]
for the evolution of the remnant in the pre-supernova wind (from
Eq.~A3 and A4 of LH03); $\nh$ is a hydrogen number density of the wind
at the radius of the blast wave ($R_b$), $E_{51} \equiv E / 10^{51}$
ergs, $E$ is the explosion energy and $M_{ej}$ is the ejected mass
from the supernova, and $M_{ej}$, $R_b$ and $\nh$ are in units of
\msun, pc and cm$^{-3}$, respectively.  
If we adopt an average radius
of $R_b$ = 7.7 pc from the geometrical center and 
the low density of $\nh$ = $0.1\, \cmthree$ that seems to be
preferred by stellar evolution model,
we obtain $t_0 = 950 M_{ej}^{3/2} E_{51}^{-1/2}$ yr and $x_0 = 6.9
M_{ej}$ pc for \gtwoninetytwo. Assuming $M_{ej}$ of several \msun\ as in
Cas A and for a canonical explosion energy of $E_{51} = 1$, we
estimate $t_0 \sim 10^4$ yr, which is larger than the remnant age ($3000$
yr) 
\citep[technically, the transition time to the ST phase ($t_{\mathrm
  ST}$) can be smaller than $t_0$, ][]{1999ApJS..120..299T}. Thus
\gtwoninetytwo\ might still be in its ED phase or in its transition to
the ST phase.
And the ED phase evolution would be better fit to
describe the current kinematics of the \gtwoninetytwo.
The evolution of the SNR in the ED phase is largely
determined by $E_{51}$, $M_{ej}$ and the power law index $n$ of the
ejecta envelope.
The radius of the blast wave ($R_b$) during the ED phase is
described by (from Eq.~A5 of LH03)
\begin{align}
\frac{R_b}{x_0} & = \left\lbrace \frac{l_{\mathrm{ED}}^{(n-2)}}{\phi_{\mathrm{ED}}}
  \frac{3}{4\pi(n-3)n} \left[ \frac{10}{3}
    \frac{(n-5)}{(n-3)}\right]^{(n-3)/2} \right\rbrace^{1/(n-2)} \notag\\
& \times \left( \frac{t}{t_0} \right)^{(n-3)/(n-2)}
\label{eq:ed}
\end{align}
where $l_{\mathrm{ED}} = 1.19 + 8/n^2$ and $\phi_{\mathrm{ED}} = 0.39
- 0.6 \exp(-n/4)$ (the physical meanings of $l_{\mathrm{ED}}$ and
$\phi_{\mathrm{ED}}$ are discussed in LH03 and TM99).  Assuming
$E_{51} = 1$, $R_b$ = 7.7 pc, $\nh$ = $0.1\, \cmthree$, and $t=3000$
yrs, the above solution gives a large ejecta mass of
$M_{ej} \sim 20\, \msun$ for the the power law index of the ejecta
envelope $n$ between 7 and 12. 
Adopting a higher wind density of $\nh$ = $0.2\, \cmthree$
reduces the ejecta mass estimate and gives $M_{ej} \sim 10 - 15\, \msun$.
The ejecta mass is also reduced if we assume a less energetic
explosion. With $E_{51} = 0.5$ and $\nh$ = $0.1\, \cmthree$, we
estimate $M_{ej} \sim 5 - 8\,\msun$.
%
These estimates are based on the assumption that the
remnant is strictly in the ED phase.
The deceleration of the remnant during its transition to the ST phase
would increase the explosion energy and decrease the ejecta mass.
%
  We note that, due to the high wind density near the center,
  radiative cooling may have been effective in the very early age of
  the remnant \citep[which may be responsible for the optical emission
  from the ejecta,][]{2003ApJ...597..347L} and this could have
  affected the overall evolution.  However, at this early time, most
  of the explosion energy would be in the form of the kinetic energy
  of the expanding ejecta, and the loss of energy by radiative cooling
  should be negligible in the total energy conservation and the
  overall remnant evolution.


The shock velocity of the blast wave ($v_b$) is given as
\[ v_b = 2500\,\kms \left(\frac{n-3}{n-2}\right)
\left(\frac{r_b}{7.7 \mathrm{pc}}\right) \left(\frac{t}{3000 \mathrm{yrs}}\right)^{-1}
\].
For the observed electron temperature range of 0.6 -- 0.9 keV, this
shock velocity indicates $\mathrm{T}_{\mathrm{e}}/\mathrm{T}_{\mathrm{p}}$ $\sim 0.1$,
consistent with those in other remnants with comparable shock velocities
\citep{2007ApJ...654L..69G}.
The optical
observations of [\ion{O}{3}] ejecta show an average expansion velocity
of $\sim 1700\,\kms$ \citep{2005ApJ...635..365G}.  This velocity may
be considered an upper limit for the current reverse shock velocity
(the velocity in the rest frame of the ambient gas;
$\frac{dR_r}{dt}$). Using Eq.~A13 of LM03,
we derive $\frac{dR_r}{dt} \simeq \frac{x_0}{t_0}$ near $t = t_0$. The
derived velocity $\frac{dR_r}{dt}$ is insensitive to $t$ because of
its logarithmic dependency. We estimate
$\frac{dR_r}{dt} \sim 2040\,\kms$  for
$E_{51} \sim 1$ and
$\frac{dR_r}{dt} \sim 1660\,\kms$ for $E_{51} \sim 0.5$, which are
consistent with the optical observations \citep{2005ApJ...635..365G}.



While we have assumed a spherically symmetric explosion, recent
observations and hydrodynamic modelings of SNe revealed a compelling case
that high-mass SN explosions are intrinsically aspherical events
\citep{2008ARA&A..46..433W,2005ASPC..332..350B,2005ASPC..332..363J}.
Previous observations of \gtwoninetytwo\ indeed suggested that the
explosion of \gtwoninetytwo\ may have been asymmetric.
\citet{2007ApJ...670L.121P} showed that the ejecta temperature is
significantly higher generally in the northwestern and western regions
of the SNR than in the southeastern regions. They suggested that this
large-scale nonuniform distribution of the ejecta temperature may be
caused by an asymmetric SN explosion: the explosion could have been
more energetic toward NW-W than in SE.
\citet{2009ApJ...692.1489W} found that [\ion{O}{3}] knots in
north--south (N--S) axis are
faster than those along the east--west axis, and suggested a
possibility that explosions were more energetic along N--S axis.
Also, there is a possibility that the azimuthal density variation
around the remnant might be related with the asymmetric explosion
(\S~\ref{sec:azim-vari-asymm}). However, these scenarios are not fully
consistent with each
other and further studies are needed.
Our future study, which will focus on the X-ray characteristics of the
shocked ejecta material, will help us reveal the detailed nature of
the explosion of \gtwoninetytwo.


\subsection{Progenitor Star}

 For the progenitor mass of $\sim 20 - 40$ \msun\ estimated
  from previous X-ray observations
  \citep{1994ApJ...422..126H,2003ApJ...583L..91G,2004ApJ...602L..33P},
  the overall observed wind properties are in plausible agreement with
  the general physical picture that the progenitor was a massive star
  which lost a significant amount of the mass through the RSG winds.
  However, given the uncertainties in the observational results and
  theoretical models, the current results do not provide a
  unique evolutionary track for the progenitor star (of a specific
  initial mass) which led to the explosion of \gtwoninetytwo. Based
  on our results, we briefly discuss a few possible scenarios for the
  evolutionary track of the progenitor.

  The progenitor star of initial mass of as low as $\sim
  20\,\msun$ seems to be consistent with lower end of our wind mass estimate,
  where 
  %
  the star had a mass loss of $\sim 10-15\,\msun$ as the RSG wind
  while the rest of the mass went into the ejecta and the neutron
  star. In this scenario, the kinematics of the remnant can be
  explained by an explosion of $E_{51}\sim0.5$. While the stellar
  evolutionary models for a single star of $M \sim 20\,\msun$ 
  seem to predict smaller mass
  losses than 10 \msun,
  a possible companion star may have
  increased the mass loss, as proposed for the case of Cas~A
  \citep{2006ApJ...640..891Y}.  A significantly increased amount of
  mass loss is expected if the progenitor is more massive.
  For example, the models suggest that a star of $30\,\msun$
  loses $\sim 20\,\msun$ as the RSG wind and than explodes as the RSG
  star \citep{2002RvMP...74.1015W}. 
  While a black hole might be expected after the explosion of such a
  massive star, the final fate of the compact remnant is
  inconclusive as observations show that a neutron star can be still
  formed from even more massive stars \citep[$> 40\
  \msun$,][]{2006ApJ...636L..41M}.  A progenitor star of $> 35\ \msun$
  likely has evolved into the WR phase after the RSG phase. This might
  be the case of \gtwoninetytwo\ if its WR phase has been short.
  In summary, given that the wind is most likely the RSG wind, the
  evolutionary models seems to prefer a progenitor's initial mass
  between $\sim20-35\ \msun$. This is plausibly consistent with the
  lower range of our wind mass estimates ($\sim 15\ \msun$),
  although the higher mass progenitor cannot be completely ruled out.


\section{Summary}
\label{sec:summary-conclusion}

Using our deep ($\sim$500 ks) Chandra observations, we present the results
from our detailed analysis of the blast wave in the Galactic O-rich
SNR \gtwoninetytwo.  
  We find that the observed emission measure of the shocked ambient
  gas increases toward the contact discontinuity while the observed gas
  temperature decreases.
  Comparisons of these observational results to self-similar solutions
  \citep{1982ApJ...258..790C} reveal that the observation is best
  described by the SNR currently expanding within the medium with a
  radially decreasing density profile ($\rho \propto r^{-2}$), likely
  a wind from the massive progenitor star.  The inferred ambient
  medium is well described by a slow RSG wind from the progenitor star
  with a mass loss rate of $\dot{M} = 2-5 \times 10^{-5}\,\msun\,
  \mathrm{yr}^{-1}$ assuming a wind velocity of $v_w \sim 10\, \kms$.
  The ambient density is too high for Wolf-Rayet wind, whose fast wind
  velocity implies a mass loss rate order-of-magnitude higher than
  what is generally accepted.  The total swept-up mass of the wind is
  estimated to be $\sim 15 - 40 \, \msun$, where the quoted range is
  due to the observed azimuthal density variation. The wind mass
  estimate may also be affected by other systematic uncertainties such
  as the distance and geometry.  Because the ambient medium is likely
  the RSG wind from a progenitor of $M \sim20-35\ \msun$, the higher
  range of our wind mass estimates ($\sim 40\, \msun$) may not be
  favored by standard stellar evolution models, although it is not
  completely ruled out.  The overall kinematics of G292.0+1.8 is
  plausibly described by the models for a remnant expanding inside the
  RSG wind.



  Our results
  provide direct observational evidence for the blast wave interacting
  with the CSM winds in \gtwoninetytwo, which is consistent with the
  core-collapse origin of this SNR as revealed based on the central
  pulsar and the O-rich nature of the metal-rich ejecta. While the
  density of the CSM wind is one of the fundamental properties to
  study the nature of young core-collapse SNRs and their
  progenitor's late-stage evolution,
  a direct observational measurement has been difficult.  Deep X-ray
  observations of young core-collapse SNRs provide an excellent
  opportunity for such a study.
  Our future studies of the nature
  of the SN ejecta in \gtwoninetytwo\ will be helpful to improve our
  understanding of the late-stage evolution of massive stars and their
  subsequent explosion as supernovae.


\acknowledgements
This work was supported in parts by the SAO under
Chandra grants G06-7049A (Penn State), G06-7049B (Johns Hopkins), and
G06-7049C (Rutgers).
B.M.G. acknowledges the support of a Federation Fellowship from the
Australian Research Council through grant FF0561298.
JPH acknowledges fruitful discussions with
Vikram Dwarkadas.


\end{document}